\documentclass{aa}
\usepackage{graphics}
\begin{document}

\thesaurus{06(08.19.5 SN 1987A; 02.14.1; 08.19.4)} 

\def\EE#1{\times 10^{#1}}
\def\gcm{\rm ~g~cm^{-3}}
\def\cm3{\rm ~cm^{-3}}
\def\kms{\rm ~km~s^{-1}}
\def\cms{\rm ~cm~s^{-1}}
\def\ergs{\rm ~erg~s^{-1}}
\def\wl{~\lambda}
\def\wll{~\lambda\lambda}
\def\Nii{M(^{56}{\rm Ni})}
\def\FeI{{\rm Fe\,I}}
\def\FeII{{\rm Fe\,II}}
\def\FeIII{{\rm Fe\,III}}
\def\Niii{M(^{57}{\rm Ni})}
\def\FeIb{{\rm [Fe\,I]}}
\def\FeIIb{{\rm [Fe\,II]}}
\def\FeIIIb{{\rm [Fe\,III]}}
\def\Msun{~{\rm M}_\odot}
\def\Ti44{M(^{44}{\rm Ti})}
\def\MZA{M_{\rm ZAMS}}
\def\mum{\mu{\rm m}}

\def\lsim{\!\!\!\phantom{\le}\smash{\buildrel{}\over
  {\lower2.5dd\hbox{$\buildrel{\lower2dd\hbox{$\displaystyle<$}}\over
                               \sim$}}}\,\,}
\def\gsim{\!\!\!\phantom{\ge}\smash{\buildrel{}\over
  {\lower2.5dd\hbox{$\buildrel{\lower2dd\hbox{$\displaystyle>$}}\over
                               \sim$}}}\,\,}

\title{ISO SWS/LWS observations of SN 1987A\thanks{ISO is an ESA project with
instruments funded by ESA Member States (especially the PI countries: France,
Germany, the Netherlands and the United Kingdom) and with the participation of 
ISAS and NASA.}}

\author{P. Lundqvist\inst{1} 
\and J. Sollerman\inst{1} 
\and C. Kozma\inst{1} 
\and B. Larsson\inst{1}
\and J. Spyromilio\inst{2}
\and A.P.S. Crotts\inst{3}
\and J. Danziger\inst{4}
\and D. Kunze\inst{5}
}
\institute{Stockholm Observatory, SE-133 36 Saltsj\"obaden, Sweden.
\and European Southern Observatory, Karl-Schwarzschild-Strasse 2, 
D-85748 Garching, Germany.
\and Columbia University, Dept. of Astronomy, 538 W. 120th Street, New
York, USA
\and Osservatorio Astronomico, Via G.B. Tiepolo 11, I-34131 Trieste,
Italy.
\and Max-Planck-Institut f\"ur extraterrestrische Physik, Postfach 1603, 
D-85740 Garching, Germany.
}

\date{Received: 03 February 1999  Accepted: 14 April 1999}

\mail{peter@astro.su.se}

\maketitle

\begin{abstract}
We report on observations of SN 1987A with ISO SWS/LWS made $9 - 11$
years after the explosion. No emission from the supernova was seen.
In particular, the upper limits on the fluxes of [Fe~I]~24.05$\mum$ and 
[Fe~II]~25.99$\mum$ on day 3\,999 are $\sim 1.1$~Jy and $\sim 1.4$~Jy, 
respectively. Assuming a homogeneous distribution of $^{44}$Ti 
inside $2\,000 \kms$, we have made theoretical models to estimate the mass of
ejected $^{44}$Ti. Assessing various uncertainties of the model, we obtain
an upper limit of $\simeq 1.5\EE{-4}\Msun$. The implications of this are 
discussed.

The LWS data display continuum emission as well as nebular lines of [O~I],
[C~II] and [O~III] from neighboring photoexcited regions in the LMC. 
The [O~III] lines indicate an electron density of $120\pm75$~cm$^{-3}$, and
the continuum can be explained by dust with a temperature of $\sim 37$ K. A 
second dust component with $\sim 10$ K may also be present.
\end{abstract}

\keywords{supernova: individual: SN 1987A -- nucleosynthesis -- supernovae: general}

\section{Introduction}

Supernova (SN) 1987A provides a convenient tool to test our understanding
of nucleosynthesis in massive stars and during supernova explosions
(Thielemann et al. 1996). In particular, 
the late ($t \gsim 150$~days) emission probes 
directly the elemental abundances deep in the stellar ejecta. As in other
core-collapse SNe without a dense circumstellar medium, the energy production
is at this epoch dominated by radioactive energy from the decay 
of $^{56}$Co to $^{56}$Fe,
the cobalt itself being the decay product of $^{56}$Ni. It was quickly realized
that the ejected mass of $^{56}$Ni in SN 1987A was at least $0.05 \Msun$ 
(Woosley et al. 1988), the actual number being close to $0.07 \Msun$
(e.g., Suntzeff \& Bouchet 1990). This decay continued to power the bolometric
light curve in an undisputed fashion for about $\sim 800 - 1\,000$ days 
(Bouchet et al. 1996).

However, during the following epoch, radioactive decay seemed to be unable
to explain the bolometric flux of the supernova, unless a large amount
of $^{57}$Ni (decaying to $^{57}$Co and further to $^{57}$Fe) was included 
(Suntzeff et al. 1992). Fransson \& Kozma (1993) demonstrated that the 
derived $^{57}$Ni/$^{56}$Ni ratio could be much closer to the 
solar $^{57}$Fe/$^{56}$Fe ratio when time dependence was accounted for.
This effect, now known as the ``freeze-out'' effect, stems from the fact
that the energy stored at earlier epochs in the 
low-density hydrogen gas slowly leaks out and dominates the bolometric light
curve after $\sim 1\,000$ days (Fransson \& Kozma 1999). The freeze-out effect
becomes less important as the radioactive isotope $^{44}$Ti starts to dominate. 
This occurs after $\sim 1\,500 - 2\,000$ days (Woosley et al. 1989;
Kumagai et al. 1991; Fransson \& Kozma 1999). $^{44}$Ti decays to $^{44}$Sc
on a time scale of $87.0\pm1.9$ years (Ahmad et al. 1998; G\"orres et al. 
1998), and then quickly further to $^{44}$Ca.

It is important to determine the $^{56}$Ni, $^{57}$Ni and $^{44}$Ti masses in
order to constrain models for the supernova explosion and the explosive
nucleosynthesis (e.g., Timmes et al. 1996). In the case of $^{44}$Ti, models
predict that $\sim 10^{-4} \Msun$ could be synthesized in SN 1987A (e.g., 
Kumagai et al. 1991; Woosley \& Hoffman 1991).

Chugai et al. (1997) estimate that the mass of ejected $^{44}$Ti should 
be $\Ti44 \sim (1-2)\EE{-4} \Msun$ from the optical line emission at 2875 days.
Kozma (1999) and Kozma \& Fransson (private communication 1999; 
hereafter KF99) 
preliminary find that $\Ti44 = (1.5 \pm 1.0) \times 10^{-4} \Msun$
best explains broad-band photometry of SN 1987A for $t \lsim 3\,270$ days.
However, nearly all of the emission comes out in the far infrared which is not
included in the observed bands. KF99 show that a more
definite estimate can be made if one could measure the
flux in a few iron lines, mainly [Fe~II]~25.99$\mum$. There are several reasons
for this. First, at $\sim 4\,000$ days KF99 find that 
almost half of the 
luminosity from the supernova is emerging in this line. Second, because the 
iron lines are only little affected by freeze-out they are good tracers of the 
instantaneous energy deposition, and the flux of [Fe~II]~25.99$\mum$ is 
therefore almost proportional to $\Ti44$. Third, this
line is formed through collisional excitations and is therefore insensitive to
uncertainties in atomic data involved in calculating the recombination cascade.
In a preliminary analysis, Borkowski et al. (1997) find that their 
Infrared Space Observatory (ISO; Kessler et al. 1996) observations give
an upper limit on the line flux which corresponds to 
only $1.5 \times 10^{-5} \Msun$. 

Here we report on observations we have made with ISO. Although we have 
observed the supernova over the entire wavelength range $2.38  - 197 \mum$, we 
will concentrate our discussion on the region, $23 - 27 \mum$, where the 
strongest emission lines are expected to emerge. We compare these observations
with theoretical modeling. We also briefly discuss our observations at longer
wavelengths where we detect emission originating from elsewhere in the LMC.

\section{Observations and Results}

We have used ISO to observe SN 1987A on several occasions.
Both the Short Wavelength Spectrograph (SWS; de Graauw et al. 1996) and the
Long Wavelength Spectrograph (LWS; Clegg et al. 1996) were used, and Table~1
summarizes our observations. We will concentrate here mainly on the SWS
observation from 4 February, 1998 (Sect. 2.1), though we provide a consistency 
check of this observation against our other SWS observations. The LWS data are
discussed in Sect. 2.2.

\begin{table}
\caption{Log of observations.}
\label{log}

\[
\begin{tabular}{lcccr}
\hline

\noalign{\smallskip}

Date & Phase$^{a}$ & Instrument/Mode & Exposure time$^{b}$ \\

\noalign{\smallskip}
\hline
\noalign{\smallskip}

96/05/26 & 3\,380 & SWS/AOT2 & 7\,307 \\
96/06/08 & 3\,393 & SWS/AOT2 & 2\,545 \\
97/07/19 & 3\,799 & SWS/AOT2 & 7\,307 \\
98/02/04 & 3\,999 & LWS/AOT1 & 3\,641 \\
98/02/04 & 3\,999 & SWS/AOT1 & 6\,750 \\

\noalign{\smallskip}
\hline
\end{tabular}
\]

$^{a}$ Epochs in days past explosion.\\
$^{b}$ Time in seconds.

\end{table}
\subsection{SWS Observations}

The SWS observation on 4 February, 1998 was made in the SWS01 mode with speed 4
and was centered on the position of SN 1987A, i.e., R.A. = 5h 35m 28.05s; 
Decl. =  $-69^{\rm o}$ 16$\arcmin$ 11$\farcs$64; (J2000.0). SWS01 provides 
spectra which together cover the entire wavelength region 
between $2.38  - 45.2 \mum$. However, as our models predict [Fe~I]~24.05$\mum$ 
and [Fe~II]~25.99$\mum$ to be by far the strongest emission lines from the
supernova at this epoch (day 3\,999; see Sect. 3), we have concentrated on 
measuring the flux in the wavelength range including these two lines. This
range, between $22.5 - 27.5 \mum$, is covered by band 3D of SWS.

The reductions were made using the SWS Interactive Analysis software system 
(SIA) available at ISO Spectrometer Data Center (ISOSDC) at the Max Planck
Institut f\"ur extraterrestrische Physik in Garching (MPE).
The most recent set of calibration files equivalent to off-line
processing (OLP/pipeline) version 7.0 was used.
The interactive reduction allows special care to be given to dark
subtractions, which is of particular interest when measuring low flux levels.
Flat fielding was also applied, but we have not made any fringe corrections 
since the fringes at low flux levels disappear in the noise.

\begin{figure}
\hskip 0.1cm \resizebox{8.5cm}{!}{\includegraphics{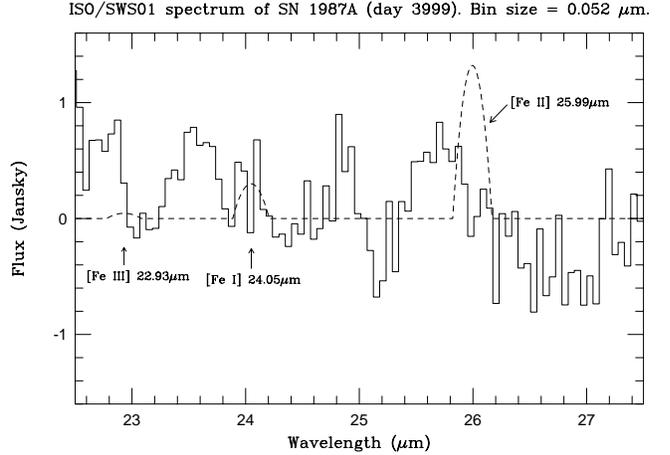}}
\caption{
  ISO SWS/AOT1 Band 3 spectrum of SN 1987A on day 3999. The
  spectrum was reduced using interactive software (see text). The bin size was
  set to $0.052 \mum$, in accordance with the instrumental resolution.
  Overlayed (dashed line) is the modeled line emission from model M2 (without
  photoionization) described in Table 2 and Sect. 3.2. To obtain the modeled
  line profiles we have assumed a maximum core velocity of $2\,000 \kms$, and 
  that the emission is homogeneously distributed throughout the core (see
  Sect. 3.3). The modeled peak flux of [Fe~II]~25.99$\mum$ corresponds 
  to what is needed for a 3$\sigma$ detection.
  }
\label{smooth.ps}
\end{figure}

In Fig. 1 we present a fully reduced spectrum of SN 1987A for band 3D.
Although the nominal instrumental resolution of SWS is $R \approx 1\,000$, the
slowest SWS01 mode with speed 4 degrades this by a factor of 2. We have
therefore averaged the spectrum with a bin size of 0.052$\mum$, 
corresponding to $600 - 650 \kms$ for 
the two lines of interest. As the lines could extend to well 
above $\pm 2\,000 \kms$ this resolution should be sufficient to resolve 
the lines. However, neither [Fe~I]~24.05$\mum$ nor [Fe~II]~25.99$\mum$ 
are seen in the spectrum. 
The `features' that do appear around $\sim 23.5$ and $25.7 \mum$ are most
likely due to instrumental effects, as they are rather robust in the sense
that they appear in many detectors and in both up and down scans. They are 
certainly not due to fringes, nor do we believe they are effects of pure noise.
The zero level of the spectrum is well defined and a simple zero-order fit to 
the spectrum gives an RMS of $\sim 0.47$~Jy over the range $22.5 - 27.5 \mum$. 

Looking more closely at the spectral regions of the two lines, we have simply
measured the RMS between $23 - 25 \mum$ and $25 - 27 \mum$ separately. Again, 
we used a zero-order fit to the smoothed spectrum. A zero-order fit should
provide the most conservative way to estimate the `noise' in our data. The 
resulting RMS values are $\sim 0.34$~Jy and $\sim 0.46$~Jy for 
[Fe~I]~24.05$\mum$ and [Fe~II]~25.99$\mum$, respectively. 
This gives 3$\sigma$ upper limits of $\sim 1.02$~Jy and $\sim 1.38$~Jy for 
the peak of the profile of the two lines. These values do not change much 
even if we double the bin size.

SN 1987A was also observed in the SWS02 mode on three occasions (see Table 1). 
This mode provides scans over shorter wavelength ranges than the full SWS01
scans. We have reduced also the SWS02 data using SIA. While [Fe~II]~25.99$\mum$
was looked at on all three occasions (see Table 1), [Fe~I]~24.05$\mum$ 
was only looked at on June 8, 1996.
The flux calibration for the SWS02 mode is in general more reliable
than for SWS01 at a specific wavelength since dark exposures are taken 
immediately before and after the short SWS02 scans, and the sampling for the
individual data points is better. However, the SWS02 observations in Table 1
cover only the range $25.75 - 26.20 \mum$, which could be just a fraction of 
the real line width. As the continuum level is therefore unknown, we have to 
rely on absolute flux calibrations, which are uncertain at the low 
flux levels of our data. Concentrating on the $26\mum$ line, we 
measure $\sim 0.75$~Jy and $\sim 0.62$~Jy for the two epochs with the longest
exposures, May 26, 1996 and July 19, 1997, respectively. The accuracy of 
absolute fluxes for this wavelength regime at higher flux levels is $< 20 \%$ 
(ISOSWS Data Users Manual, v5.0), but for our very low flux 
levels, $\sim 50 \%$ is probably more fair. 

Due to this uncertainty we can only say that our SWS02 flux estimates are 
consistent with the upper limits we obtain from the SWS01 measurement.

\subsection{LWS Observations}

Our LWS observations were made on 4 February, 1998. The full range of 
the LWS was scanned, i.e., $45 - 197 \mum$. Standard reduction of the spectra
using the ISO Spectral Analysis Package (ISAP) software resulted in the 
combined spectrum displayed in Fig. 2. Just as for the SWS observations, 
no emission was observed from the supernova. 
However, four unresolved emission lines from neighboring
photoexcited gas were identified: [O~I] 63.2$\mum$, [O~III] 51.8$\mum$, 
[O~III] 88.4$\mum$ and [C~II] 158$\mum$, 
with the fluxes $(1.5 \pm 0.3) \times 10^{-12}$ erg s$^{-1}$ cm$^{-2}$, 
$(1.39 \pm 0.13) \times 10^{-11}$ erg s$^{-1}$ cm$^{-2}$, 
$(1.68 \pm 0.32) \times 10^{-11}$ erg s$^{-1}$ cm$^{-2}$, and
$(3.36 \pm 0.12) \times 10^{-12}$ erg s$^{-1}$ cm$^{-2}$, respectively.
The lines sit on top of a broad continuum which peaks at $\sim 100 \mum$ at
a level of $(17 \pm 5)$~Jy, and which is only marginally weaker in the 
range $\sim 145-165 \mum$.
The integrated continuum flux is $(6 \pm 2) \times 10^{-10}$ erg s$^{-1}$
cm$^{-2}$, which for a distance of 50 kpc corresponds to a luminosity 
of $(1.8 \pm 0.6) \times 10^{38}$ erg. It should be pointed out that the 
spatial region sampled has a large diameter ($\sim 80 \arcsec$), which at a 
distance of 50~kpc distance corresponds to $\sim 19$~pc. We discuss the LWS 
results briefly in Sect. 4.3.

\begin{figure}
\hskip 0.1cm \resizebox{8.5cm}{!}{\includegraphics{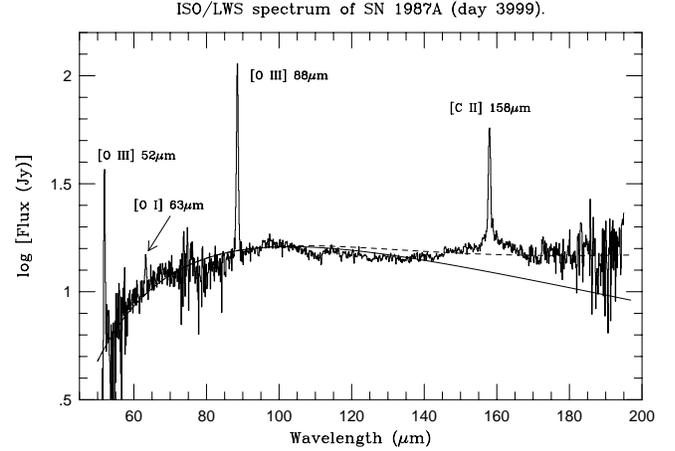}}
\caption{
  ISO LWS spectrum of SN 1987A on day 3999. The continuum up to 140$\mum$ can 
  be fitted with a spectrum emitted by dust with a temperature of $\sim 37$ K 
  (solid smooth line). Longward of 140$\mum$ a second component appears to 
  add in. The temperature characterizing this component is $\sim 10$ K. The
  combined emission is shown by the dashed line. See Sect. 2.2 and 4.3 for 
  details.
  }
\label{lws.ps}
\end{figure}

\section{Interpreting the SWS observations}

To understand what the upper limits on the iron lines in Sect. 2.1 mean, we 
have made model calculations. The models are described in Sect. 3.1, and we
discuss the results from the calculations in Sect. 3.2

\subsection{The model for line emission from the supernova}
The code we have used for our theoretical calculations 
is described in detail in Kozma \& Fransson (1998a), and
here we will only give a brief summary of the model.

The thermal and ionization balances are solved time-dependently, 
as are also the level populations of the most important ions.

The ions included are H I-II, He I-III, C I-III, N I-II, O I-III,
Ne I-II, Na I-II, Mg I-III, Si I-III, S I-III, Ar I-II, Ca I-III,
Fe I-V, Co II, Ni I-II. 
The following ions are treated as multilevel atoms: H I, He I, O I, Mg I, 
Si I, Ca II, Fe I, Fe II, Fe III, Fe IV, Co II, Ni I, Ni II. The remaining
lines are solved as 2-level atoms or as $4-6$ level systems.
For Fe I we include 121 levels,
Fe II 191 levels, Fe III 112 levels and for Fe IV 45 levels.
A total of approximately 6\,400 lines are included in the calculations.
 
The radioactive isoptopes included are $^{56}$Ni, $^{57}$Ni, and $^{44}$Ti.
These radioactive decays provide the energy source for the ejecta, and we 
calculate the energy deposition of gamma-rays and positrons solving the 
Spencer-Fano equation (see Kozma \& Fransson [1992] for a more detailed 
description of our thermalization of the gamma-rays). We have assumed that the 
positrons deposit their energy locally, within the regions containing the 
newly synthesized iron.

We have included $0.07 \Msun$ of $^{56}$Ni (Sect. 1)
and $3.33\EE{-3} \Msun $ of $^{57}$Ni, which corresponds to a 
$^{57}$Fe/$^{56}$Fe ratio equal to two times the solar ratio. This mass 
of $^{57}$Ni gives a good fit to the bolometric light curve when the effects 
of freeze-out are taken into account (Fransson \& Kozma 
1993). It is also consistent with observations of [Fe II] and [Co II] IR 
lines (Varani et al. 1990; Danziger et al. 1991), observations of the $^{57}$Co
122 keV line (Kurfess et al. 1992), and theoretical nucleosynthesis 
calculations (Woosley \& Hoffman 1991). In our `standard' model we have 
used $1.0\EE{-4} \Msun$ of $^{44}$Ti (e.g., Kumagai et al. 1991; Timmes et al.
1996). We have also made calculations with $\Ti44 =0.5\EE{-4} \Msun$ 
and $\Ti44 = 2.0\EE{-4} \Msun$.

Until recently, the lifetime of $^{44}$Ti has been very poorly known. 
In our calculations we have used an e-folding time of 78 years thought
to be a representative value to those found from experiments. However, while 
completing our calculations, we were informed about the recent measurements 
mentioned in Sect. 1 (S. Nagataki, private communication). The new,
and accurate value, $87.0\pm1.9$ years, is close to what we have used, but
decreases the instantaneous radioactive energy input from the $^{44}$Ti decay
by $\lsim 12\%$. This systematic error, shifting our estimated value of $\Ti44$
upward by a similar amount, has been included in the error analysis 
in Sect. 4.1.

As input for the calculations we have adopted the abundances from the 10H 
explosion model (Woosley \& Weaver 1986; Woosley 1988). Spherically symmetric
geometry has been assumed, with shells containing the major composition 
regions as given by the explosion model. We have used the same distribution
of shells as in Kozma \& Fransson (1998a) where hydrogen is mixed
into the core. The iron-rich core is assumed to extend out to $2\,000 \kms$, 
outside of which we attach a hydrogen envelope, out to $6\,000 \kms$.

For the line transfer we have used the Sobolev approximation.
This is justified for well-separated lines in an expanding medium.
However, especially in the UV, there are many overlapping lines,
and one can expect UV-scattering to be important. The importance of 
line scattering decreases with time as the optical depths decrease.
The effect of line scattering is to alter the emergent UV-spectrum,
but it also affects the UV-field within the ejecta. The ionization of
elements with low ionization potential is sensitive
the UV-field. During scattering the UV photons are 
shifted towards longer wavelengths, both due to a pure Doppler shift, but
also because of the increased probability of splitting the UV-photons into 
several photons of longer wavelengths. A more accurate treatment of the line
scattering is therefore expected to decrease the importance of
photoionization. For the modeling of [Fe~II]~25.99$\mum$, the photoionization
of Fe~I to Fe~II is likely to introduce the largest uncertainty. To check the
importance of this effect we have in some models simply switched off the
photoionization. A more comprehensive line transfer modeling is discussed in 
KF99.

\subsection{Model calculations}
The results of our calculations can be seen in Table 2. We tabulate line 
fluxes for four strong lines for three values of $\Ti44$: $5\EE{-5}$ (M1),
$1\EE{-4}$ (M2) and $2\EE{-4}$ $\Msun$ (M3). 
For each value of $\Ti44$ we show
results for models with and without photoionization. In all six models
a simple form of dust absorption was assumed (Kozma \& Fransson 1998b). The 
line fluxes in Table 2 are for a box line profile between $\pm 2\,000 \kms$,
and assuming a distance to the supernova of 50 kpc. 

From Table 2 it is evident that [Fe~II] 25.99$\mum$ is by far the strongest
line in our simulations, and that its flux scales roughly linearly 
with $\Ti44$, as expected from Sect. 1. From power-law fits to the results
of the calculations described in Table 2, 
we obtain $f_{26\mum} \propto \Ti44^{1.13}$ when photoionization 
is included, and nearly the same without photoionization 
($f_{26\mum} \propto \Ti44^{1.12}$). The second strongest line 
is [Fe~I] 24.05$\mum$. Its dependence on $\Ti44$ is 
weaker: $f_{24\mum} \propto \Ti44^{0.56}$ 
and $f_{24\mum} \propto \Ti44^{0.73}$, respectively. Other lines from SN 1987A
are far too weak for our ISO observations.

The stronger dependence of $f_{26\mum}$ on $\Ti44$ than for $f_{24\mum}$ is
due to ionization: while a higher $\Ti44$ boosts the relative fraction of 
Fe~II, $X_{\rm FeII}$, the relative fraction of Fe~I, $X_{\rm FeI}$, 
decreases. For example, in models with photoionization, 
the average value of $X_{\rm FeII}$ in the Fe-rich gas 
increases from $\sim 0.32$ to $\sim 0.55$ when $\Ti44$ is increased 
from $5\EE{-5}$ to $2\EE{-4}$, while $X_{\rm FeI}$ decreases from $\sim 0.65$ 
to $\sim 0.36$. The fact that the 26$\mum$ line is so much stronger than 
the 24$\mum$ line despite the roughly similar relative fractions of Fe~I 
and Fe~II is just a result of the much larger collision strength of the 
Fe~II line than for [Fe~I] 24.05$\mum$.

Table 2 shows that switching off photoionization does not have a dramatic
effect on the line fluxes. In the case of M2, $f_{26\mum}$ decreases 
by $\sim 15 \%$ and $f_{24\mum}$ increases by $\sim 28 \%$ when we turn off 
photoionization, simply due to a minor shift in ionization between Fe~I and 
Fe~II. As we will discuss in Sect. 4.1, there are other uncertainties which
are of the same magnitiude.

\begin{table}
\caption[]{Modeled line flux in Jy at 4\,000~days$^{a,b}$}
\label{modelfluxes}
\[
\begin{tabular}{lccc}
\hline
\noalign{\smallskip}

Line & M1$^{c}$ & M2$^{d}$ & M3$^{e}$ \\

\noalign{\smallskip}
\hline
\noalign{\smallskip}


$\FeIb~24.047\mum$   & 0.11 (0.12) & 0.16 (0.20) & 0.23 (0.33) \\
$\FeIIb~25.995\mum$  & 0.45 (0.40) & 1.03 (0.88) & 2.19 (1.89) \\
$\FeIIb~35.359\mum$  & 0.01 (0.01) & 0.04 (0.04) & 0.12 (0.11) \\
$\FeIIIb~22.930\mum$ & 0.01 (0.01) & 0.03 (0.03) & 0.11 (0.11) \\

\noalign{\smallskip}
\hline
\end{tabular}
\]

$^{a}$ Distance = 50 kpc. Box line profile of width $\pm 2\,000 \kms$.\\
$^{b}$ For values in parantheses photoionization has been excluded.\\
$^{c}$ $\Ti44 = 5\EE{-5}\Msun$\\
$^{d}$ $\Ti44 = 1\EE{-4}\Msun$\\
$^{e}$ $\Ti44 = 2\EE{-4}\Msun$\\

\end{table}

\subsection{Mass of $^{44}$Ti from [Fe~II] 25.99$\mum$}
With our results it is straightforward to estimate the upper limit on $\Ti44$. 
Using $f_{26 \mum} \lsim 1.38$~Jy from Sect. 2.1, and our power-law fit to the 
results in Table 2 (Sect. 3.2), we find an upper limit
which is $\Ti44 \lsim 1.3 (1.5) \times 10^{-4} \Msun$ with (without)
photoionization included. These masses are for a box-shaped line profile,
and therefore provide conservative limits. However, there is good reason to
believe that the line profile should be similar to those for the lines observed
by Haas et al. (1990) $\sim 400$ days after the explosion. The strongest line 
observed by Haas et al. was [Fe~II] 17.94$\mum$ and it had 
$v_{\rm FWHM} = (2\,900 \pm 80) \kms$. Although it peaked just short off 
$+ 1\,000 \kms$, and thus was not symmetric around the rest velocity of the
supernova, it had the general appearance of the expected line profile formed
by a filled sphere with constant emission throughout the sphere. 
For a sphere extending out to $2\,000 \kms$, the peak is 1.5 
times higher than for the box profile used in Table 2 
(which is valid for a hollow sphere), and $v_{\rm FWHM} = 2\,828 \kms$. 
Using such a line profile in our upper limits on $\Ti44$, 
we instead obtain $\Ti44 \lsim 0.9 (1.0) \times 10^{-4} \Msun$ with (without)
photoionization included. Within the framework of our modeling, 
a conservative limit (i.e., the case when photoionization is unimportant) for 
a plausible line profile is therefore $\Ti44 \lsim 1.0 \times 10^{-4} \Msun$.
In Fig. 1, we have included the expected line emission for such a model 
with this limiting mass of $^{44}$Ti. We will evaluate this limit on $\Ti44$ 
in Sect. 4.1.

\section{Discussion}

\subsection{Uncertainties in the modeling}
There are several uncertainties involved in our modeling of the line fluxes. 
We have already checked the effect of the lifetime of $^{44}$Ti (Sect. 3.1).
We have also studied the effect of photoionization, and found that it
introduces rather mild uncertainties. A similar level of uncertainty is due
to the distance to the supernova. This is still inaccurate to the level 
of $5 - 10 \%$ (Lundqvist \& Sonneborn 1999; Walker 1999), which means an 
uncertainty in the line flux of $\sim 10 - 20 \%$.

Atomic data of iron are notoriously difficult to calculate accurately.
This is therefore another source of error in our modeling, especially for
individual lines. Most important for the $26 \mum$ line is the collision 
strength of that transition, $\Omega_{26 \mum}$. One normally assigns an 
uncertainty in the collision strength for the strongest iron lines 
to $\sim 30\%$ (e.g., Verner et al. 1999). In our models we have 
used $\Omega_{26 \mum} = 5.8$ (Zhang \& Pradhan 1995). To study the effect 
in detail we have tested a model with $\Ti44 = 10^{-4} \Msun$ 
and $\Omega_{26 \mum} = 2.9$. We find that there is not a linear scaling
between $f_{26 \mum}$ and $\Omega_{26 \mum}$, as one might naively believe. 
Instead, we find from linear interpolation that a $30\%$ decrease in 
collision strength gives a $\sim 17\%$ lower $f_{26 \mum}$, and a 
correspondingly higher estimate of $\Ti44$. The reason for this is that the 
gas is slightly hotter in the model with reduced collision strength, 
boosting the exponential term in the collisional rate so that it somewhat 
counteracts the reduced collision strength. We have recently found out 
(A. Pradhan, private communication) that the preferred value
for $\Omega_{26 \mum}$ at the low temperatures in SN 1987A is most 
likely closer to $\sim 7.0$, i.e., higher than we have used. Despite this, 
we have generously assigned an uncertainty of $15\%$ (upwards) in $\Ti44$ 
due to atomic data.

Another source of uncertainty could be the explosion model used.
In these calculations we have used the abundances from the 10H explosion model. 
A comparison between the two models 10H (Woosley \& Weaver 1986; Woosley 1988)
and 11E1 (Shigeyama et al. 1988) was done in Kozma \& Fransson
(1998b). There it was found that the iron lines are not sensitive to the 
explosion model used, because the iron core mass is the same
in both models. The iron core mass is set by the amount of $^{56}$Ni which
is accurately determined from the bolometric light curve. The choice of 
explosion model thus does not seem to be a major source of uncertainty when
modeling these iron lines.

In our calculations we assume a local deposition of the positrons originating
from the radioactive decays of $\Ti44$. We believe this is a good
approximation since optical and near-IR light curves of Fe~I and 
Fe~II lines show that trapping must occur (Chugai et al. 1997; 
Kozma \& Fransson 1998b). Actually, there is no obvious sign of a leakage
of positrons, neither from broad-band lightcurves (KF99), nor from the optical
Fe I lines at $6\,300$ \AA\ until the last data point at $3\,597$ days in
Kozma (1999). Although the trapping may well be fully complete, we have 
assigned an error to this assumption by 15\%.

Another approximation in our models is the assumption of a homogeneous
density in each Fe-rich shell of the model core. To test the sensitivity to 
this assumption, we have run a model similar to M2, with photoionization 
included (see Table 2), but where we have divided the mass in the Fe-rich
ejecta into two components of equal mass but with different densities.
The denser component is set to be nearly five times more dense than the other.
Despite the significantly different density distribution in this model compared
to that in M2, the differences in $f_{26 \mum}$ and $f_{24 \mum}$ between
the models are small. Instead of the values listed in Table 2, the fluxes in 
the two-component model are 0.98 and 0.20 Jy, respecively, i.e., $f_{26 \mum}$ 
differs by only $\sim 5\%$ compared to that in M2. This indicates that the
model assumption of a homogeneous density in each Fe-rich shell does not 
introduce a major uncertainty.

Finally, screening and cooling by dust are potential sources of error 
in our models. The effects of dust are examined in Kozma \& Fransson (1998ab).
The screening we use is discussed in Kozma \& Fransson (1998b), and is based on
estimates by Lucy et al. (1991) and Wooden et al. (1993). Dust from pure iron 
is unlikely to form, as that would cool and block out all iron line emission 
once the dust has formed. On the contrary, there is a wealth of iron lines from
the core at late times. KF99 and Kozma (1999) find good agreement between 
modeled and observed broad-band lightcurves for $t \lsim 3\,270$ days and 
spectra at $2\,870$ days. As was pointed out also for the positron leakage, 
the models are also able to reproduce optical Fe I lines at even later epochs.
Based on this, we believe dust effects are small enough to neglect in our 
estimate of $\Ti44$.

None of the model approximations we have used appears to be uncertain enough to 
allow $f_{24 \mum}$ to be of the same magnitude as $f_{26 \mum}$. The best 
estimate of $\Ti44$ (within the framework of our modeling) should therefore 
come solely from $f_{26 \mum}$. To estimate the combined error of $f_{26 \mum}$ 
due to all model approximations (except for the dust distribution in the 
ejecta), we adopt the uncertainties $15\%$, $15\%$, $15\%$ and $5\%$ for 
photoionization, distance, atomica data and clumping, respectively. For 
the choice of input model and positron leakage we adopt $15\%$ each. This gives
a combined uncertainty which is $\sim 34\%$. On top of this we must add the
maximum systematic error of $12\%$ discussed in Sect. 3.1 for the lifetime 
of $^{44}$Ti. With the line profile discussed in Sect. 3.3, we therefore
arrive at an upper limit on $\Ti44$ which is $\simeq 1.5 \EE{-4} \Msun$.
 We note that this limit excludes the
upper ends of the allowed ranges of $\Ti44$ found by Chugai et al. (1997) and 
KF99 (see Sect. 1). Combining our limit with the preliminary results 
of KF99 for the broad-band photometry, a likely range 
for $\Ti44$ is $(0.5 - 1.5)\EE{-4} \Msun$. We emphasis, however, that 
the lower limit of this range is probably more uncertain than the upper 
(for the reasons mentioned in Sect. 1), which is indeed indicated by the
preliminary analysis of Borkowski et al. (1997).

\subsection{Implications of the derived mass of $^{44}$Ti}

Models for the yield of $^{44}$Ti give quite different results. This is most 
likely due to how the explosion is generated in the models, and how fallback
onto the neutron star is treated. Timmes et al. (1996; see also Woosley \&
Weaver 1995) use a piston to generate the explosion, 
and they account for fallback in a rather self-consistent way. In their
model with zero-age mass $\MZA = 20 \Msun$ (i.e., corresponding to SN 1987A)
the mass of the initially ejected $^{44}$Ti is $1.2\EE{-4} \Msun$, but 
of this only $1.4\EE{-5} \Msun$ escapes after fallback. This is less than we 
argue for in Sect. 4.1, and could suggest that fallback was not important 
for SN 1987A, though we caution again that the lower limit found by KF99
(see also Kozma 1999) may not be very strict. If fallback is unimportant the
ejected amounts of $^{56}$Ni and $^{57}$Ni would be too high in this model,
typically by a factors of $\sim 2-4$, judging from the effects of fallback in
the $25 \Msun$ model in Woosley \& Weaver (1995). It should be emphasized
that the variation of $\Ti44$ with $\MZA$ in Timmes et al. (1996) 
is complex, and that for models with $\MZA = 18 \Msun$ 
and $\MZA = 22 \Msun$, the calculated $\Ti44$ comes within the range we 
propose, albeit close to our lower limit.

The models of Thielemann et al. (1996) simulate the
explosion by depositing thermal energy in the core, and they insert the mass
cut artificially so that the right amount of ejected $^{56}$Ni is produced.
(This effectively means that fallback is included also in these models.)
Simulating the explosion in this way ensures larger entropy and thus more
alpha-rich freeze-out than in Woosley \& Weaver (1995). Accordingly, the
ratio $\Ti44 / \Nii$ (where $\Nii$ is the mass of ejected $^{56}$Ni that
does not fall back) is higher in the models of Thielemann et al. than
in piston-driven simulations. For example, in the $20 \Msun$ model of 
Thielemann et al. (1996) $\Nii \approx 0.074 \Msun$, 
$\Niii \approx 2.9\EE{-3} \Msun$ and $\Ti44 \approx 1.7\EE{-4} \Msun$, 
with $\Niii$ defined in the same way as $\Nii$ and $\Ti44$. The values
of $\Nii$ and $\Niii$ are close to what have been inferred for SN 1987A
(Suntzeff \& Bouchet 1990; Fransson \& Kozma 1993). The titanium mass is 
slightly larger than the upper limit of the range we estimate in Sect. 4.1.
So, while our estimate of $\Ti44$ cannot rule out with certainty any of 
the two methods used for the explosion (piston-driven or heat generated), 
our results could indicate that an intermediate method 
should be used (see also the discussion on this in Timmes et al. 1996).
From models of the chemical evolution of the Galaxy, and especially the solar 
abundance of $^{44}$Ca, a value for $\Ti44$ closer to that of Thielemann et 
al. (1996) may be more correct, at least for supernovae in general.

In this context we note that a higher value of $\Ti44$ is produced in
asymmetric explosions (Nagataki et al. 1998). The method of calculation 
employed by Nagataki et al. (see Nagataki et al. 1997) is similar to that in 
Thielemann et al. (1996), though the models of Nagataki et al. allow for
2-D instead of just 1-D. With no asymmetry, the models of Nagataki et al.
produce $\Ti44 \sim 6\EE{-5} \Msun$ for an explosion similar to SN 1987A, 
when the mass cut has been trimmed to $\Nii \approx 0.07 \Msun$. This is 
significantly less than Thielemann et al. (1996) despite the similar method 
of modeling. Applying an asymmetry by a factor of 2 between the equator and
the poles, the explosion energy becomes concentrated toward the equator 
resulting in relatively stronger alpha-rich freeze-out, which increases $\Ti44$
to $\sim 2\EE{-4} \Msun$ (for $\Nii \approx 0.07 \Msun$). This is outside 
our observationally determined range and could indicate that asymmetry was not
extreme in SN 1987A, especially if $\Ti44$ is close to the limit derived
by Borkowski et al. (1997). We note that a piston-driven calculation could 
perhaps allow for asymmetry, as such models give very small values of $\Ti44$
in 1-D.

A direct way to estimate $\Ti44$ in supernovae is to observe the gamma-ray 
emission from the radioactive decay of $^{44}$Ti. The 1.156 MeV line associated 
with the decay of $^{44}$Ti has only been detected in two supernova remnants
(and no supernovae): Cas A (Iyudin et al. 1994; The et al. 1996) and the newly
discovered gamma-ray/X-ray source GRO J0852-4262/RX J0852.0-4622 (Iyudin et al.
1998; Aschenbach 1998). While $\Ti44$ in the latter is not yet 
well-known, $\Ti44$ in Cas A was $1.7^{+ 0.6}_{-0.5}\EE{-4} \Msun$
(G\"orres et al. 1998). In the models of Nagataki et al. (1998), the value
of $\Ti44$ at the upper end of this range would indicate that Cas A exploded
asymetrically. Our estimate of $\Ti44$ in Sect. 4.1 does not seem to
indicate a high degree of asymmetry for SN 1987A, but an independent 
measurement of its 1.156 MeV line may be needed to be more conclusive
on this point. To detect this emission from SN 1987A, instruments like 
INTEGRAL (Leising 1994) are required.

\subsection{Interpretation of LWS observations}

The two [O~III] lines observed (see Sect. 2.2 and Fig. 2) can be used to 
estimate the average density of the 
emitting gas. We have used a multi-level atom to do this, using the atomic data
of Mendoza (1983), Osterbrock (1989) and Aggarwal (1993). 
Assuming a temperature of $10^4$ K, we obtain a mean electron density 
of $120 \pm 75 \cm3$ in the [O~III] emitting gas. 

The continuum can be fitted with a spectrum of the 
form $F_{\lambda} \propto B_{\lambda}(T) (1 - {\rm exp}(-\tau_{\rm D}))$, 
where $\tau_{\rm D}$ is the dust optical depth. With a functional form 
of $\tau_{\rm D} = \tau_{0.55} (0.55 / \lambda)^{\alpha}$ (where $\lambda$ is 
in microns) we obtain a best fit with $T \simeq 37$ K, $\tau_{0.55} \simeq
0.74$ and $\alpha \simeq 0.83$. The fit (see Fig. 2) works well up 
to  $\simeq 140 \mum$, where an extra source appears to add in.
This could indicate the presence of cold dust. Assuming the same
functional form for this emission as for the $37$ K component, 
the temperature of the cold component is close to 10 K.

\section{Conclusions}
We have looked at SN 1987A with ISO SWS/LWS. The supernova was not 
detected in any of the spectra. In particular, we have derived upper limits 
for the fluxes of [Fe~I]~24.05$\mum$ and [Fe~II]~25.99$\mum$, and made time
dependent calculations for the late line emission from the supernova. We have
assessed various uncertainties to the models, and for a plausible line
profile we then find an upper limit on the mass of ejected $^{44}$Ti,
$\Ti44 \simeq 1.5\EE{-4} \Msun$. Together with the preliminary results 
of KF99 this brackets the ejected mass of $^{44}$Ti to 
the range $\Ti44 = (0.5 - 1.5)\EE{-4} \Msun$, which is close to the yield 
of $^{44}$Ti in models of the explosion by Woosley \& Weaver (1995), 
Thielemann et al. (1996) and Nagataki et al. (1997). The lower limit
is probably less stringent than the upper, as indicated by the results of
Borkowski et al. (1997). A more direct limit on $\Ti44$ can be made when 
instruments measuring the gamma-ray line emission from the $^{44}$Ti decay 
will be available.

The only emission our ISO observations detect is from gas and dust in 
the direction toward the supernova. In particular, the [O~III] 51.8$\mum$
and [O~III] 88.4$\mum$ lines indicate a gas density of $120\pm75$~cm$^{-3}$, 
and the dust continuum can be explained by a temperature of $\sim 37$ K. 
A second dust component with $\sim 10$ K may also be present.

\begin{acknowledgements}

We thank Claes Fransson for comments and stimulating discussions, and Bruno
Leibundgut and the referee Tino Oliva for comments on the manuscript. 
We also thank Shigehiro Nagataki and Anil Pradhan for updating us on
nuclear and atomic data, respectively. We acknowledge support from the 
ISOSDC at MPE. ISAP is a joint development by the LWS and SWS Instrument Teams
and Data Centers. Contributing institutes are CESR, IAS, IPAC, MPE, RAL and 
SRON. This research was supported by the Swedish Natural Science Research 
Council, the Swedish National Space Board, and the Knut \& Alice Wallenberg 
Foundation. A.P.S.C. is grateful for support from NASA grants NAG5-3319 and 
NAG5-3502.

\end{acknowledgements}

{}

\end{document}